\documentclass[a4paper]{cas-dc}

\usepackage[authoryear,longnamesfirst]{natbib}
\usepackage{times} 
\usepackage[T1]{fontenc}
\usepackage{mathptmx}

\usepackage{graphicx}
\def\tsc#1{\csdef{#1}{\textsc{\lowercase{#1}}\xspace}}
\tsc{WGM}
\tsc{QE}
\tsc{EP}
\tsc{PMS}
\tsc{BEC}
\tsc{DE}

\begin{document}
\let\WriteBookmarks\relax
\def\floatpagepagefraction{1}
\def\textpagefraction{.001}
\shorttitle{}
\let\printorcid\relax

\title [mode = title]{Impact of Short-Duration Aerobic Exercise Intensity on Executive Function and Sleep}

\author[1]{Yu Peng}
\author[1]{Guoqing Zhang}
\cormark[1]
\ead{Pengyuzhang9999@163.com}

\author[2]{Huadong Pang}

\address[1]{Department of Physical Education, Hunan Institute of Science and Technology, Yueyang, 414000, China}
\address[2]{Georgia Institute of Technology, Atlanta, 30332, USA}

\begin{abstract}
IoT-based devices and wearable sensors are now common in daily life, with smartwatches, smartphones, and other digital tools tracking physical activity and health data. This lifelogging process provides valuable insights into people's lives. This paper analyzes a publicly available lifelog dataset of 14 individuals to explore how exercise affects mood and, in turn, executive function. Results show that moderate physical activity significantly improves mood, reduces stress, and enhances cognitive functions like decision-making and focus. Improved mood not only boosts exercise performance but also strengthens executive function, suggesting exercise benefits both emotional and cognitive well-being. This opens the door for personalized exercise plans tailored to emotional states to optimize brain function.

\end{abstract}



\maketitle

\section{Introduction}

\begin{figure}[h]
	\centering
	\includegraphics[width=0.5\textwidth]{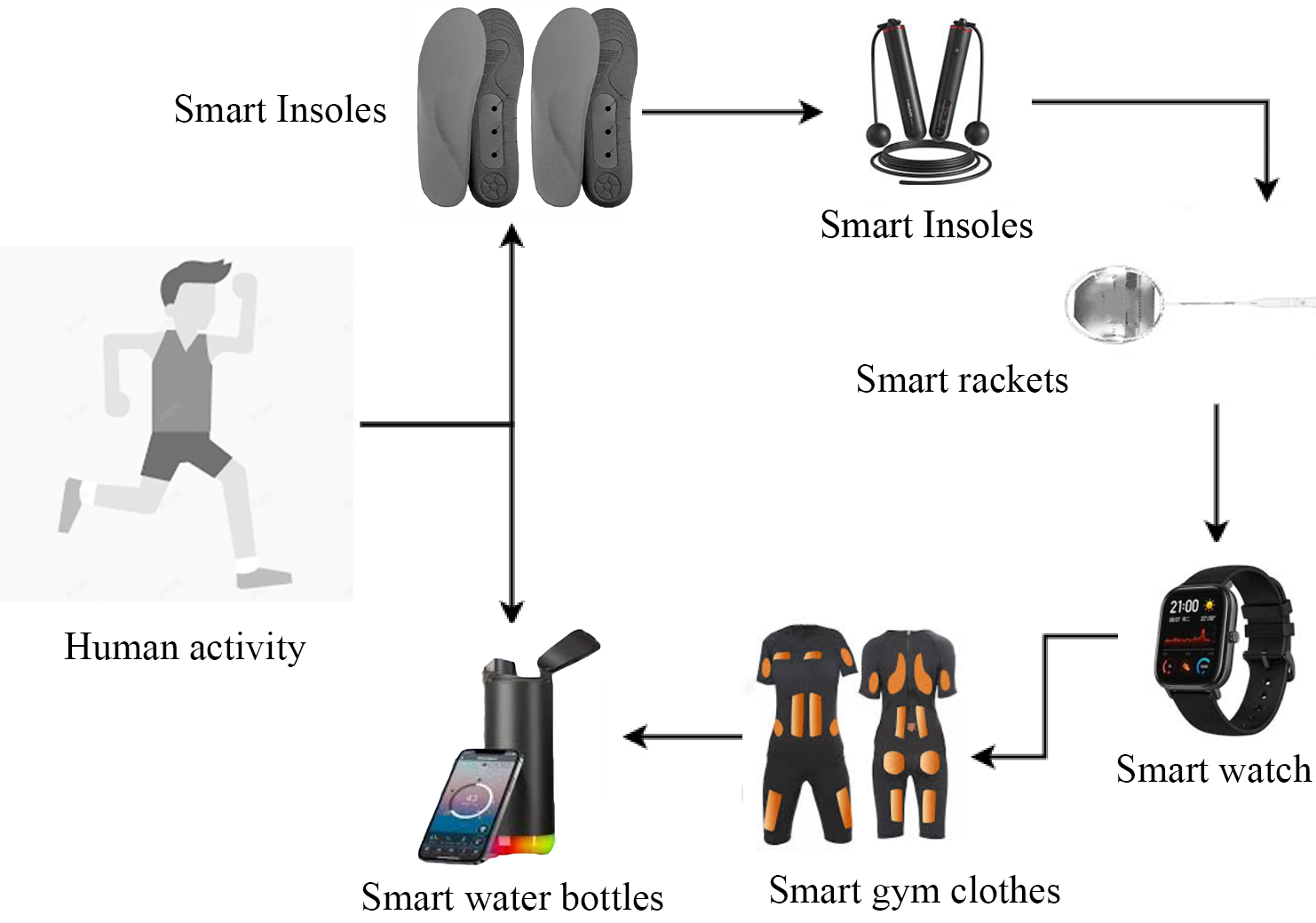}
        \caption{Local Self-Attention Block}
	\label{IoT_Figure}
\end{figure}

With the proliferation of Internet of Things (IoT) devices and wearable sensors, a vast amount of physical activity and mental state data is accurately recorded in what is known as lifelogging \cite{mehren2019intensity}. Long-term, effective exercise not only improves individual health but can also reveal fluctuations in mood and psychological states through data collected by IoT devices. Scientifically designed exercise programs, combined with data monitoring from wearable devices, can enhance both physical health and athletic performance \cite{ludyga2016acute,tsai2021acute}. Figure \ref{IoT_Figure} illustrates the types of wearable device technologies utilized. These wearables, such as smartwatches, insoles, and body sensors, can be affixed to the bodies of middle-aged individuals to detect physiological parameters such as temperature, heart rate, and blood pressure while also recording their physical activity data\cite{ishihara2021effects,chen2014effects}. Users can access the recorded information via cloud platforms, facilitating quality-of-life analysis. Specifically designed for each individual, these wearable devices can track movement vectors within a virtual environment. They are equipped to evaluate local positioning, with a primary objective of enhancing walk prediction systems to improve health monitoring efficacy\cite{coetsee2017effect,dupuy2018effect,,YU2024111200,dupuy2018effect}. The advantages of wearable and IoT technologies in the context of task-based health management underscore their effectiveness in assisting users in documenting their activity levels. Through real-time data analytics, these devices not only track the physical activities of middle-aged individuals but also monitor physiological metrics, thereby promoting the maintenance of a healthy lifestyle and enabling the timely identification of potential health issues.

The relationship between emotions and physical activity is becoming a focal point of research. Increasing evidence suggests that emotions, such as stress, fatigue, and pleasure, not only affect athletic performance but also alter an individual’s motivation and execution through complex psychological mechanisms\cite{moreau2019acute,yuxin2021effects}. Physiological metrics captured by IoT devices, such as heart rate, step count, and calorie expenditure, combined with emotional state data (e.g., self-reported mood and stress levels), can reveal deeper connections between emotions and physical activity. Studies have shown that positive emotional states enhance exercise performance, while negative emotions, such as anxiety or fatigue, may reduce physical efficiency\cite{huang2022effects,park2022effects}.

To improve the design and deployment of IoT-based health monitoring systems for more accurate tracking of emotional and stress levels, several key advancements can be made. First, the diversity and precision of sensors need to be enhanced. Current IoT devices typically rely on physiological data such as heart rate and step count to infer emotional or stress levels, but these metrics do not directly measure emotional fluctuations. To address this limitation, future systems could integrate sensors that directly capture emotional states, such as Galvanic Skin Response (GSR), Electroencephalography (EEG), and Heart Rate Variability (HRV) sensors. These sensors provide real-time insights into the autonomic nervous system activity, brain wave patterns, and emotional states, enabling a more accurate assessment of emotions and stress. The integration of these advanced sensors will offer more precise and multidimensional data, improving the system's ability to evaluate and respond to an individual’s emotional condition.

In addition to sensor upgrades, data fusion and algorithm optimization~\cite{peng2024automatic,wang2024deep,Wang2024Theoretical,li2024ltpnet,wan2024image,richardson2024reinforcement} are crucial for enhancing the accuracy of emotional and stress level monitoring. By integrating data from multiple sensors, such as heart rate, GSR, and EEG, into a unified analytical framework, the system can provide a comprehensive understanding of an individual’s emotional state. Machine learning and deep learning algorithms can further enhance the system's ability to predict and identify emotional changes based on physiological data. Additionally, incorporating edge computing technology can allow for real-time processing and feedback, ensuring that the system responds promptly to emotional fluctuations. Real-time adjustments to personalized health interventions, such as modifying exercise intensity or providing relaxation suggestions, can help users manage their emotional states effectively. By leveraging these technologies, future IoT health monitoring systems can offer highly personalized and accurate emotional and stress level tracking, improving both physical and mental well-being.

However, the impact of emotions is not limited to physical activity; it is also closely associated with executive functions. Executive functions refer to the ability to manage, regulate, and coordinate cognitive resources in complex tasks, encompassing areas such as inhibitory control, task switching, and working memory. Research indicates that emotions can significantly affect various dimensions of executive function. Stress and negative emotions, such as anxiety or frustration, can impair inhibitory control, leading to impulsive decision-making or difficulties in maintaining attention\cite{soga2015executive,soga2015executive}. By analyzing physiological data and emotional states captured by IoT devices, researchers can further investigate how emotions influence physical performance through their effect on executive functions, and conversely, how physical activity may enhance executive functions by improving emotional regulation\cite{tsukamoto2016greater,alves2012effects,aguirre2019effect}.

This offers a novel perspective for the development of personalized health and fitness recommendation systems. By integrating physical activity data with emotional states, it is possible not only to enhance the understanding of individual athletic performance but also to predict the potential impact of emotional states on executive functions, thereby aiding in the creation of more targeted health interventions and exercise plans\cite{wang2013executive}. Ultimately, such systems would be able to comprehensively consider both physical and mental states, providing more personalized recommendations to help users achieve better physical health and cognitive performance.

\subsection{Life log data and athletic performance}

The collection of lifelogging data is derived from a wide range of sources, relying on various smart devices, including smartphones, wearable devices (such as smartwatches and health trackers), computers, smart glasses, digital cameras, and other smart wearables\cite{hsieh2021systematic,hsieh2021systematic}. These devices can collect, analyze, and transmit real-time information about an individual’s physical condition, daily activities, and surrounding environment, generating rich data streams. Additionally, social media platforms provide a unique perspective for collecting lifelogging data. For instance, content posted on social media, such as food photos, allows researchers to infer users' dietary habits (e.g., vegetarian, vegan, or non-vegetarian) and estimate calorie intake\cite{kamijo2009acute,park2019beneficial}. The widespread application of lifelogging data has sparked interest not only among general users, enabling them to better organize and manage daily life, but also across various research fields. In sports, medicine, and business, researchers extract key insights by mining lifelogging data\cite{wang2023effects,song2019effects}. These data can be utilized to develop personalized health and fitness recommendation systems. For example, monitoring athletes’ physical activity, heart rate, food intake, and environmental conditions can aid in analyzing their recovery levels during rehabilitation \cite{varela2012effects}. By integrating and analyzing such data, researchers can further advance personalized health management, making users’ exercise and health decisions more scientifically informed and efficient.

The richness of life logging data~\cite{ren2025iot,wang2024using,liu2025real,wang2024cross,zhuang2020music,peng2025integrating} enables researchers to analyze and assess athletic performance from multiple perspectives. Depending on the type of physical activity, different metrics are often used to evaluate performance. For example, in jump-based sports, performance can be assessed by measuring the athlete's jump height and its variation over time, which reflects their progress or the effectiveness of training\cite{tari2021exercise,heath2018post,,NING2024102033,labelle2013decline}. In endurance activities, such as aerobic fitness, performance can be evaluated by observing an individual’s walking or running speed within a set time period. For instance, in one study, researchers used the six-minute walk test to assess participants' aerobic fitness by analyzing their walking speed during this time, revealing their cardiorespiratory endurance\cite{paschen2019effects,naderi2019effects}. Although the definition of athletic performance varies across different sports, commonly used key metrics include movement speed, average heart rate, calorie expenditure, and weight changes\cite{alshurideh2022components}. These indicators not only reflect an individual’s performance over time but also serve as important references for monitoring health and fitness improvements. By integrating lifelogging data collected through Internet of Things (IoT) devices and wearable sensors, researchers can more comprehensively evaluate athletic performance and explore its complex relationships with emotional states, mental health, and lifestyle factors. This provides strong support for personalized health and fitness recommendation systems and opens new pathways for improving personal athletic performance and health management.

\subsection{Relationship between exercise and mood and executive function}

\begin{figure}[h]
	\centering
	\includegraphics[width=0.5\textwidth]{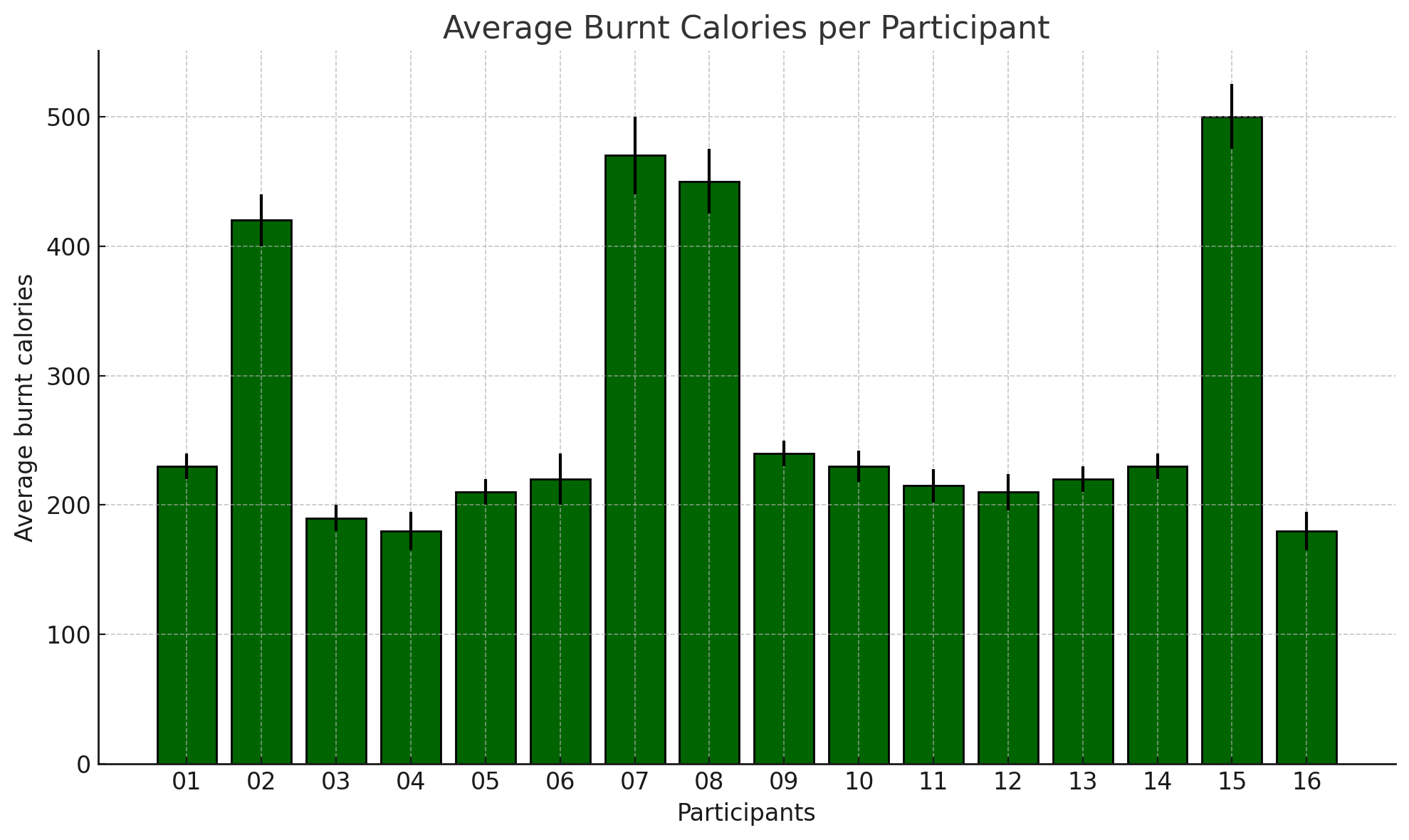}
        \caption{Local Self-Attention Block}
	\label{Figure3}
\end{figure}

The relationship between physical activity and emotions has been the subject of extensive examination in scientific literature, with an increasing body of evidence indicating that regular exercise has a marked beneficial impact on emotional regulation. The practice of engaging in physical activity has been demonstrated to result in the discharge of endorphins and other biochemical substances throughout the body. These substances have the capacity to improve one's emotional state, mitigate the effects of stress, and even serve to alleviate symptoms of anxiety and depression. Furthermore, research suggests that aerobic exercise and strength training have additional benefits in terms of emotional stability, stress reduction, and anxiety alleviation. \cite{mittelstadt2023influence}.The emotional advantages of exercise have been demonstrated to enhance overall mental wellbeing in daily life, with the potential to improve performance on tasks that require executive function. The term "executive functions" is used to describe a set of higher-order cognitive abilities that are essential for the completion of complex tasks. Such functions include inhibitory control, task switching, and working memory.
Emotions are of pivotal importance with respect to the functioning of the executive system. It has been demonstrated in numerous studies that positive emotional states facilitate cognitive control, thereby enabling individuals to display enhanced flexibility and efficiency when attempting complex tasks.\cite{clayson2024registered}. Conversely, negative emotions, such as anxiety or stress, may impair inhibitory control and attention, reducing the speed and accuracy of task switching. The mechanisms through which emotions influence executive function are complex and multifaceted, but considering the positive effects of exercise on mood, it can be hypothesized that regular physical activity may indirectly enhance executive function by improving emotional states. This interaction between emotions and executive function offers a new perspective on how exercise not only benefits physical health but also indirectly boosts cognitive and decision-making abilities through emotional regulation. As shown in Figure \ref{Figure3}, the number of steps and distance correlate most strongly with calories burned, while speed does not show a significant correlation with calories burned but has a high correlation with distance. These findings are consistent with previous studies \cite{yeung2023changes} , suggesting that this dataset could be used to explore the impact of physical activity on emotions and their influence on executive function performance.

This study makes significant contributions to the understanding of the impact of short-term aerobic exercise intensity on executive function, particularly in middle-aged individuals, addressing several gaps in the current literature. The specific contributions are as follows:

\begin{enumerate}
    \item Detailed Analysis of the Impact of Aerobic Exercise Intensity on Executive Function: By categorizing aerobic exercise intensity into low, moderate, and high levels, and considering different sleep quality conditions, this study systematically evaluates the effects of aerobic exercise intensity on various aspects of executive function, including inhibition, shifting, and updating. This detailed design reveals the differential effects of exercise intensity on executive functions, providing strong support for future cognitive interventions based on exercise.
    
    \item Addressing the Research Gap on the Combined Effects of Sleep and Aerobic Exercise on Executive Function: While existing studies typically focus on the isolated effects of sleep or exercise on executive function, this research is the first to explore the combined impact of sleep quality and aerobic exercise intensity on executive function. This novel approach provides a new perspective on the complex interaction between sleep, exercise, and executive function, offering valuable insights for future research in this area.
    
    \item Comprehensive Analysis Using Multi-Source Data: The study leverages the PMData dataset, which includes both physiological metrics (such as heart rate and exercise levels) and subjective health data (such as fatigue, stress, and mood). By integrating these diverse data sources, the study provides a holistic view of how aerobic exercise influences emotional fluctuations and executive function, offering valuable insights for personalized health management and the development of smart health technologies.
\end{enumerate}

The rest of the paper is organized as follows: Section II provides a brief review of related works in the field of Wearable sensors and Psychological well-beings and physiological features. Section III details our research design and method. This section includes sections on the dataset, sleep and aerobic exercise regimens, experimental procedures, and statistics and analysis methods. Section IV presents the experimental results of our method on datasets, and Section V concludes the paper.

\section{Background and related work}

\subsection{Wearable sensors and exercise data}
Traditionally, personal physiological data has been recorded using expensive professional and medical equipment, which often requires expert operation and comes at a high cost\cite{liu2020effects}. However, over the past decade, we have witnessed the rapid development and widespread adoption of affordable wearable sensors, such as smartwatches and fitness trackers. These devices can effectively and continuously monitor an individual's health status and physiological metrics. Compared to traditional equipment, these wearable devices are more cost-effective and can easily integrate with existing smart devices, such as smartphones and tablets, to comprehensively manage and analyze daily health data\cite{cui2022associations}].

Scientific research has demonstrated that physiological data captured by wearable sensors holds significant value across various fields, not only limited to health management and fitness applications but also in clinical monitoring and disease prevention. The data obtained from these sensors has been successfully applied in numerous health management scenarios, including health monitoring, activity tracking, and performance optimization\cite{langenberg2022repeated}. For instance, in the study conducted by\cite{van2013structure,spaniol2022meta}, the authors evaluated heart rate monitoring across four different types of devices: professional sports equipment with chest straps, fitness trackers, smartwatches, and smartphones. The results showed that, during mild physical activity, the heart rate measurements from fitness trackers and smartwatches were highly accurate. However, during high-intensity exercise or when heart rate fluctuated significantly, discrepancies in measurements were observed across the devices. The study also demonstrated that heart rate patterns could be used for activity recognition and to intelligently recommend new exercises based on the user’s individual experience, current physiological state, and progress.

Additionally, the study in\cite{hillman2009effect} investigated the effectiveness of using digital health trackers to observe the relationship between weight fluctuations and activity tracking. The findings revealed a positive correlation between activity tracking and weight loss, particularly among participants who frequently logged their food intake, as they showed greater weight loss. Similarly, those who exercised regularly achieved more significant weight reduction compared to those who exercised less frequently. These findings suggest that wearable sensors have great potential and practical value in helping users manage their weight, improve their health, and create personalized health plans.

\subsection{Psychological well-beings and physiological features}

An individual's health encompasses both physical and mental well-being \cite{vandendaele2023lexical}. The findings of numerous studies indicate that exercise has a beneficial effect on the physiological and psychological functions of individuals undergoing various forms of treatment. For example, in study \cite{langenberg2023understanding}, The authors conducted an examination of the effects of aerobic exercise on the physiological characteristics and psychological well-being of cancer survivors. The findings indicated that low-to-moderate-intensity aerobic exercise programs may prove an efficacious means of improving the physical and mental health of those affected by cancer, offering a valuable approach to rehabilitation. This research underscores the significance of structured exercise in promoting recovery. In addition to enhancing physical strength, exercise has been demonstrated to facilitate improvements in mood and emotional stress, underscoring the multifaceted benefits of structured exercise for cancer survivors.

In study \cite{cheng2023effects}, The relationship between the physiological and psychological variables was subjected to further investigation. The researchers conducted a three-month study with 17 participants, collecting comprehensive life-logging data to analyze the interrelationship between physical and mental health. A range of physiological variables was captured using wireless sensors, encompassing activity levels, bed occupancy, nocturnal heart rate and respiration, daytime heart rate, blood pressure, step count, body weight, and environmental factors like room lighting and temperature. In addition to the aforementioned measurements, the participants provided a daily self-assessment of their stress levels, mood, and sleep quality. Behavioral data, including information on exercise and sleep patterns, was collected using mobile diaries.

The findings indicated a notable correlation between stress levels and a range of physiological variables. For example, elevated stress levels were frequently linked to elevated heart rate and blood pressure, suboptimal sleep quality, and irregular activity patterns. These findings indicate a strong correlation between an individual's psychological state and their physiological responses. For instance, chronic stress may manifest as changes in vital signs, sleep disruptions, and alterations in daily activity. By monitoring these variables concurrently, researchers were able to obtain a more profound understanding of the interrelationship between physical and mental health in real time.

This study highlights the potential for utilizing wearable sensors and self-reported data to comprehensively monitor an individual's physical and mental health. Through the continuous capture of physiological data and its correlation with subjective psychological assessments, healthcare providers may gain a more nuanced understanding of an individual's overall well-being. Furthermore, this approach could facilitate the implementation of more personalized interventions, wherein exercise programs, stress management techniques, and lifestyle modifications are tailored to address both the physical and emotional needs of patients. The combination of physiological and psychological data in the context of health monitoring presents new avenues for holistic health management and rehabilitation. This integration enables the development of more efficacious strategies for enhancing overall health outcomes.

\subsection{Lifelog data and exercise performance modeling}
In the context of using lifelog data for analyzing sports and exercise performance, study \cite{guimond2008psychological} explored the use of heart rate and GPS data in conjunction with computational intelligence~\cite{liu2024dsem,zhao2025short,pang2024electronic,huang2024risk,yan2024application,an2023runtime,guo2024construction} to analyze cycling and other physical activities.By means of comprehensive data analysis, it was possible for the researchers to ascertain a positive correlation between heart rate and altitude gradient. This indicates that when athletes engage in activities at elevated altitudes or on inclines, their heart rates increase. However, a negative correlation between heart rate and speed was also identified, indicating a decline in heart rate with increasing speeds. It is possible that this finding may be influenced by a number of factors, including the terrain, the level of fitness or endurance of the athlete in question, and other external or individual factors. It should therefore be considered with caution. These observations illustrate the intrinsic difficulty of interpreting physiological data when considering diverse environmental conditions. The study underscored the potential of computational intelligence as an indispensable instrument for the manipulation and examination of intricate physiological and geographical data.

Similarly, study\cite{netz2007effect} focused on predicting athlete performance by analyzing physical features and using Support Vector Machines (SVM) in combination with swarm optimization techniques. SVM is a highly resilient machine learning algorithm with the capacity to effectively process high-dimensional data and identify optimal classification boundaries. The authors utilized a dataset comprising 500 records pertaining to 100-meter sprint events. Despite the restricted sample size, the findings indicated that the SVM model demonstrated enhanced performance in the prediction of athletic performance compared to models based on linear regression and neural networks. Although linear regression models are often effective in predicting linear relationships and neural networks demonstrate proficiency in handling intricate nonlinear challenges, the SVM exhibited remarkable generalization capabilities, even with restricted data. Consequently, the SVM demonstrates considerable potential for utilisation within the domain of sports analytics.

More recently, study \cite{sibley2006effects} further advanced this field by utilizing the Predictive Power Score (PPS) of lifelog data features to build sports performance prediction models. PPS is a technique employed to evaluate the predictive relationships between features, thereby enabling researchers to identify hitherto obscured correlations within the lifelog data. By employing the technique of partial proportionality scaling (PPS) analysis, the authors were able to discern a multitude of relationships patterns, including those of a non-linear nature, asymmetric correlations, and predictive values between categorical and nominal data. In contrast with the methods traditionally employed in correlation analysis, PPS is capable of identifying intricate and interactive relationships, thereby affording researchers a more profound understanding of feature selection. In light of the aforementioned selected features, a series of predictive models was then developed, comprising linear regression models, multilayer perceptron (MLP) networks, convolutional neural networks (CNN), and long short-term memory (LSTM) networks~\cite{jin2025rankflow}. It is important to note that each of these models possesses distinctive strengths. The models based on linear regression are well suited to the prediction of simple linear relationships. Meanwhile, the aforementioned models based on multilayer perceptron (MLP), convolutional neural network (CNN), and long short-term memory (LSTM) demonstrate enhanced capacity for addressing intricate nonlinear relationships and time series data.

Notably, the authors argued that PPS is a highly effective feature selection approach, particularly in the development of sports performance prediction models. Through this method, researchers can identify the most predictive features from large datasets, thereby improving the accuracy and efficiency of the models. However, despite the significant findings, the predictive models in study\cite{pahor2022near} did not include some key wellness attributes, such as fatigue, readiness, soreness, stress, and mood. These factors play a crucial role in an athlete's overall performance, and their exclusion might limit the completeness of the prediction models. In the future, incorporating these wellness variables could further enhance model accuracy and provide athletes with more comprehensive performance predictions and health management insights.

\section{Research design and method}

\subsection{Dataset}

\begin{figure}[h]
	\centering
	\includegraphics[width=0.5\textwidth]{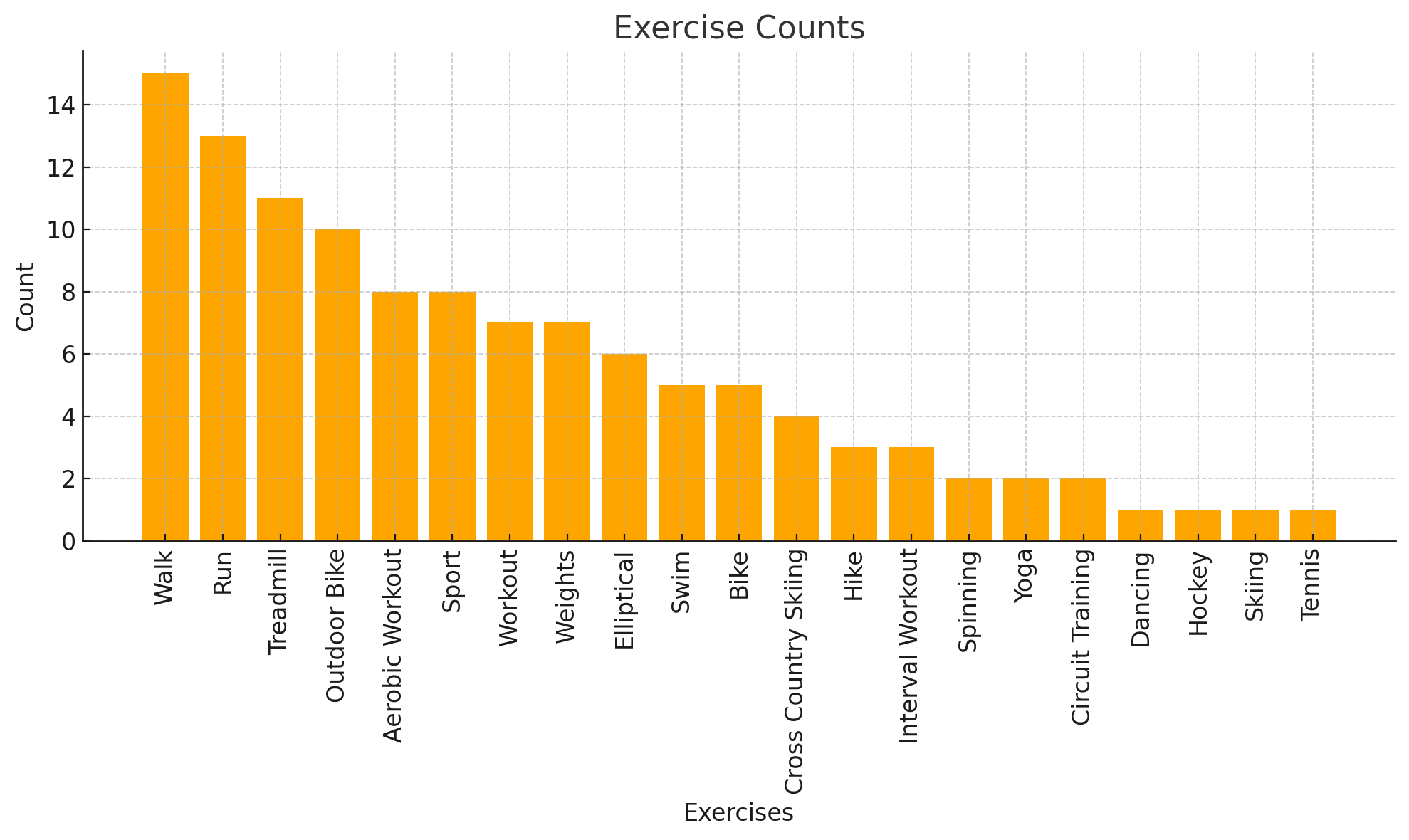}
        \caption{Local Self-Attention Block}
	\label{Figure2}
\end{figure}

The PMData dataset\cite{thambawita2020pmdata} aggregates lifelogging and physical activity data recorded through various devices and methods, aimed at facilitating in-depth analysis of athletic performance and its related factors. The data, collected over five months from 16 participants using Fitbit Versa 2 smartwatches, the PMSys sports logging application, and Google Forms surveys, encompasses both objective physiological metrics (e.g., heart rate, exercise duration, calorie expenditure) and subjective health assessments (e.g., fatigue, stress, mood). These data serve as a robust foundation for developing machine learning models that explore the relationship between lifelogging and physical activity, particularly in predicting athletic performance, weight fluctuations, and sleep patterns. As illustrated in Figure \ref{Figure2}, the most frequently performed activities among participants were "walking," followed by "running" and "treadmill," while the least tracked were "skiing" and "tennis." Additionally, participants showed a preference for regular fitness activities such as "outdoor cycling," "aerobic exercise," and "weight training," over more specialized sports like "skiing" and "hockey." 

The structure of this dataset comprises various types of data~\cite{zhou2024optimization,zhang2024deep,jiang2020dualvd,liu2025eitnet,lyu2024optimized,xi2024enhancing}, including detailed physiological metrics recorded by smartwatches and subjective health status reported through the PMSys application. Specifically, the Fitbit data~\cite{xu2022dpmpc,wang2024intelligent,yuan2025gta,tang2025real,wang2024recording,lee2024traffic} includes daily step counts, exercise distances, heart rate, and sleep information, while the subjective data from PMSys encompasses parameters such as fatigue, mood, readiness, stress, and soreness, aiding researchers in assessing participants' psychological and physiological conditions. Additionally, Google Forms collected data on dietary intake and weight changes, which are linked to participants' exercise behaviors and subjective health states. All data are synchronized via timestamps, allowing researchers to analyze the interrelationships between different data sources.

The PMData dataset integrates multiple data sources, including both detailed physiological measurements from smartwatches and subjective health reports from the PMSys application. Fitbit data provides daily metrics such as step count, distance traveled, heart rate, and sleep patterns, while PMSys reports include subjective parameters like fatigue, mood, readiness, stress, and soreness, offering a comprehensive view of participants' mental and physical states. Additionally, dietary intake and weight change data collected via Google Forms are correlated with participants' exercise behaviors and subjective well-being. All data points are time-synchronized, allowing for a detailed analysis of the relationships between the various data streams.

Although the PMData dataset provides extensive data on physiological activity and subjective health states, including heart rate, fatigue, mood, stress, and sleep quality, it lacks direct measurements related to executive functions. Executive functions are typically assessed through specific cognitive tasks, such as inhibitory control, task switching, and working memory, which are crucial for a comprehensive understanding of the relationship between emotion and cognitive function. Extensive research has demonstrated a strong link between emotions and executive functions, with stress, fatigue, and mood fluctuations significantly impacting inhibitory control, task switching, and decision-making abilities. Therefore, this study will infer the potential impact of mood fluctuations on executive functions based on conclusions from existing literature, and indirectly analyze this relationship using the emotional indicators (e.g., mood, stress, and fatigue) available in the dataset. This approach compensates for the absence of direct executive function data in the dataset and provides a theoretical basis for future research.

\subsection{Sleep and aerobic exercise regimens}

We recorded and analyzed the participants' sleep quality based on the PMData dataset. This dataset categorizes participants' sleep duration into two groups: 6-7 hours and 7-8 hours. These data provide us with insights into various sleep parameters, including total sleep duration, sleep stages, and nocturnal disturbances. By comparing these different sleep durations, we can conduct a detailed analysis of how varying intensities of short-duration aerobic exercise may interact with and influence executive function in relation to sleep quality. This approach enhances the rigor of our study, making the data analysis more systematic and comprehensive.

 The establishment of aerobic exercise intensity was guided by the American College of Sports Medicine's recommendations for aerobic intensity levels suitable for healthy adults\cite{alshurideh2022components}. Specifically, low-intensity aerobic exercise was defined as 50\% to 59\% of the individual's maximum heart rate, moderate-intensity as 60\% to 69\%, and high-intensity as 70\% to 79\%. The maximum heart rate was calculated using the formula: 220 minus age. Throughout the cycling exercises, the average heart rates achieved by participants across the three intensity levels were 53\% of maximum heart rate for low intensity, 66\% for moderate intensity, and 75\% for high intensity.

The aerobic exercise protocol was conducted using a MONARK 834 model ergometer (manufactured in Sweden), with resistance levels adjusted according to the designated exercise intensity and participant requirements, ranging from 0 to 150 watts. The cycling cadence was maintained at a minimum of 30 revolutions per minute (r/min). To monitor heart rate during the aerobic exercise at varying intensity loads, a RS800CXSD heart rate telemetry device (manufactured in Finland) was employed; timing for each exercise session commenced once participants reached target heart rate zones corresponding to the prescribed intensity levels, with a total exercise duration of 30 minutes per session.

\subsection{Experimental Procedures}

The experiment was conducted at 8:00 AM and comprised four main components: a baseline assessment, a low-intensity aerobic exercise segment, a moderate-intensity aerobic exercise segment, and a high-intensity aerobic exercise segment. In the baseline assessment, participants rested for 10 minutes before measuring their resting heart rate, followed by the administration of three cognitive tasks designed to evaluate executive function. During the low-intensity aerobic exercise segment, participants engaged in 30 minutes of low-intensity cycling on a stationary ergometer. After completing the exercise, they rested until their heart rate returned to within ±10\% of baseline levels, at which point they performed the cognitive tasks again to assess executive function. The moderate and high-intensity segments mirrored the low-intensity protocol, varying only in the exercise intensity. The order of participation in the four segments utilized a Latin square counterbalancing design to mitigate potential practice effects. To account for circadian rhythms and reduce exercise fatigue, participants completed each segment on different days, with a minimum interval of seven days between sessions. Additionally, the sequence of the three executive function tasks was randomized using a counterbalanced design to eliminate order effects. Throughout the experiment, each participant underwent three interventions of varying intensities (low, moderate, and high) and completed executive function assessments at four points (baseline, after low-intensity, after moderate-intensity, and after high-intensity exercise). An illustrative example of the experimental procedure for a given participant is presented in Table \ref{Table_1}.

\begin{table}[ht]
\centering
\caption{Experimental Procedure Example}
\resizebox{\linewidth}{!}{
\label{Table_1}
\begin{tabular}{ccc}
\hline
Time Flow & Experimental Procedure & Aerobic Exercise Plan \\ \hline
0 & Baseline Section & None \\ 
$\geq 7$ Days & Low-Intensity Aerobic Exercise & 30 minutes of low-intensity exercise \\ 
$\geq 14$ Days & Moderate-Intensity Aerobic Exercise & 30 minutes of moderate-intensity exercise \\ 
$\geq 21$ Days & High-Intensity Aerobic Exercise & 30 minutes of high-intensity exercise \\ \hline
\end{tabular}
}
\end{table}

All participants engaged in the aerobic exercise interventions and executive function tasks in a quiet, spacious, and well-lit laboratory environment. Each participant was individually assessed by a primary examiner in a one-on-one format. Prior to each cognitive task, the examiner facilitated practice trials to ensure that participants thoroughly understood the task requirements, only commencing the formal assessments once proficiency was demonstrated. The total duration of the executive function measurement was approximately 30 minutes, during which participants were allowed brief intervals of rest between each task to optimize performance and minimize fatigue.

\subsection{Statistics and analysis}

Data results were expressed as mean values ± standard deviation (M±SD) and analyzed using the Statistical Package for the Social Sciences (SPSS) version 17.0. A multivariate repeated measures analysis of variance (ANOVA)\cite{langenberg2022repeated,langenberg2023understanding} was employed to assess the effects of intensity factors, gender factors, and their interaction on executive functions, specifically inhibition, shifting, and updating. In instances where significant effects were identified, post hoc comparisons were conducted using the Bonferroni method to account for multiple comparisons, with statistical significance set at p < 0.05.
\section{Result}

We present the results of our analyses on the effects of sleep quality and short-term aerobic exercise intensity on various components of executive function, including inhibition, shifting, and updating. The focus is on gender differences and how these factors interact in middle-aged individuals and college students. We conducted repeated measures ANOVA to assess the impact of different sleep durations (6–7 hours vs. 7–8 hours), exercise intensities (baseline, low, moderate, and high), and gender (male vs. female) on executive functions. The results are discussed in terms of their significance and implications for improving cognitive performance through aerobic exercise, with detailed analysis of each executive function component. Tables summarizing the data for each function are provided to illustrate the differences in performance across conditions.

\subsection{Impact of sleep quality and short-term aerobic exercise intensity on executive function: gender differences}

A repeated measures ANOVA was conducted with a 2 (sleep quality: 6–7 hours, 7–8 hours) × 4 (intensity: baseline, low, moderate, and high) × 2 (gender: male, female) design to investigate the effects of varying intensities of short-duration aerobic exercise and gender on overall executive function in college students, encompassing inhibition, shifting, and updating. The results indicated a significant main effect of intensity within the group, F(9, 20) = 13.95, p < 0.01. However, the main effect of gender between groups was not significant, F(3, 26) = 0.86, p > 0.05, nor was the interaction effect between gender and intensity, F(9, 20) = 0.30, p > 0.05. These findings suggest that different intensities of short-duration aerobic exercise have a significant impact on overall executive function, which does not vary with changes in gender.

\subsection{Effects of sleep quality and short-term aerobic exercise intensity on inhibitory control: gender differences}

The results measuring inhibitory function in middle-aged individuals following baseline, low-intensity, moderate-intensity, and high-intensity short-duration aerobic exercise, under varying sleep conditions, are presented in Table 2. A multivariate repeated measures analysis of variance was conducted to assess the effects of different intensities of short-duration aerobic exercise on inhibitory function. Participants were categorized into two groups based on sleep quality, with one group averaging 6 to 7 hours of sleep, and the other averaging 7 to 8 hours. The analysis revealed a significant intensity effect, F(3,84) = 2.76, p < 0.05, indicating that different intensities of short-duration aerobic exercise exert varied effects on inhibitory function. Post-hoc comparisons indicated that there were no significant changes in inhibitory function following low-intensity aerobic exercise compared to baseline (p > 0.05). In contrast, both moderate and high-intensity aerobic exercises significantly enhanced inhibitory function (ps < 0.05), without significant differences between the two higher intensity levels (p > 0.05), suggesting that improving inhibitory function in middle-aged individuals requires at least moderate-intensity short-duration aerobic exercise (see Table 2). Furthermore, the main effect of gender was found to be non-significant, F(1,28) = 2.68, p > 0.05, indicating no substantial differences in inhibitory function between genders; additionally, the interaction between intensity and gender was also non-significant, F(3,84) = 0.55, p > 0.05, suggesting that the impact of different intensities of short-duration aerobic exercise on inhibitory function does not vary by gender.

\begin{table*}[htbp]\tiny%
\centering
\caption{Measurement Results of Inhibitory Function in Middle-aged Individuals After Different Sleep Durations and Different Intensity Short-term Aerobic Exercises ($M\pm SD$, unit: ms)}
\resizebox{\linewidth}{!}{
\begin{tabular}{lccccc}
\hline
\textit{Sleep Duration} & {Gender} & {Baseline} & {Low Intensity} & {Moderate Intensity} & {High Intensity} \\ \hline
 6 - 7h & Male & 8.90±12.86 & 8.70±15.74 & -1.69±21.00 & -3.84±21.30 \\ \hline
 6 - 7h & Female & 13.43±9.85 & 8.92±24.30 & 5.70±11.33 & 6.62±12.34 \\
 7 - 8h & Male & 9.0±13.86 & 8.98±13.74 & -1.53±21.00 & -3.76±21.30 \\
 7 - 8h & Female & 14.67±8.5 & 9.34±24.30 & 5.89±11.33 & 7.32±12.34 \\ \hline
\end{tabular}
}
\end{table*}

\subsection{Impact of sleep quality and aerobic exercise intensity on updating function: gender differences}
The results of measuring cognitive refresh functions after short-term aerobic exercise at baseline, low intensity, medium intensity, and high intensity among university students with varying sleep quality are presented in Table 3.

A repeated measures ANOVA was conducted to analyze the effects of different sleep qualities and varying intensities of short-term aerobic exercise on refresh functions. The analysis revealed a significant main effect of intensity, F(3,84)=18.78, p<0.01, indicating that short-term aerobic exercise of different intensities had a distinct impact on refresh functions. Post hoc comparisons showed that refresh functions significantly improved with low, medium, and high-intensity exercise compared to the baseline (ps<0.05). Moreover, refresh functions were significantly better after low-intensity exercise compared to high-intensity exercise (p<0.05), while no significant difference was found between medium and high intensities (p>0.05). These results suggest that short-term aerobic exercise enhances refresh functions regardless of sleep quality, with the best results observed for students who had longer sleep duration and performed low-intensity exercise (see Table 3). The gender effect was not significant, F(1,28)=0.08, p>0.05, indicating no notable differences in refresh functions between males and females across different sleep qualities. Additionally, the interaction between intensity and gender was not significant, F(3,84)=1.53, p>0.05, showing that the effect of different intensities of short-term aerobic exercise on refresh functions did not vary by gender.

\begin{table*}[htbp]\tiny%
\centering
\caption{Measurement Results of Inhibitory Function in Middle-Aged Adults After Different Sleep Qualities and Aerobic Exercise of Varying Intensities (M±SD, Unit: ms)}
\resizebox{\linewidth}{!}{
\begin{tabular}{lccccc}
\hline
 Sleep Duration & {Gender} & {Baseline} & {Low Intensity} & {Moderate Intensity} & {High Intensity} \\ \hline
6 - 7h & Male  & 865.96±183.79 & 737.82±154.49 & 838.8±196.19 & 781.76±176.72 \\ \hline
6 - 7h & Female  & 865.96±183.79 & 737.82±154.49 & 838.8±196.19 & 781.76±176.72 \\
7 - 8h & Male  & 935.68±192.51 & 734.89±124.18 & 822.99±115.53 & 792.77±154.9 \\
7 - 8h & Female  & 902.34±186.23 & 737.34±138.29 & 829.39±159.36 & 785.78±164.34 \\ \hline
\end{tabular}
}
\end{table*}

\subsection{Effects of sleep quality and aerobic exercise intensity on shifting function: gender differences}

The measurement results of cognitive flexibility in university students under varying sleep quality conditions after baseline, low-intensity, moderate-intensity, and high-intensity short-term aerobic exercise are presented in Table 4. A multivariate repeated measures ANOVA was conducted to examine the effects of different intensities of short-term aerobic exercise on cognitive flexibility. The analysis revealed a significant effect of exercise intensity, F(3,84)=18.25, p<0.01, indicating differences in cognitive flexibility across varying intensities. Post-hoc comparisons showed significant improvements in cognitive flexibility for low, moderate, and high-intensity exercises compared to baseline (p<0.01). However, no significant differences were found between low and moderate intensities when compared to high intensity (ps>0.05), or between low and moderate intensities themselves (p>0.05), suggesting that all levels of short-term aerobic exercise enhanced cognitive flexibility without significant differences. Gender did not have a significant effect on cognitive flexibility, F(1,28)=0.15, p>0.05, and the interaction between intensity and gender was also non-significant, F(3,84)=0.46, p>0.05, indicating that the impact of different intensities of short-term aerobic exercise on cognitive flexibility was consistent across genders, regardless of sleep quality.

\begin{table*}[htbp]
\centering
\caption{Measurement Results of Switching Function in Middle-aged Individuals after Short-term Aerobic Exercise of Different Intensities and Sleep Qualities (M±SD, Unit: ms)}
\resizebox{\linewidth}{!}{
\begin{tabular}{lccccc}
\hline
Sleep Duration & {Gender} & {Baseline} & {Low Intensity} & {Moderate Intensity} & {High Intensity} \\ \hline
6 - 7h & {Male}  & 315.33±89.79       & 223.61±80.76         & 223.09±69.23         & 213.21±72.14         \\ \hline
6 - 7h & {Female}  & 312.4±91.85        & 222.37±53.17         & 242.54±100.65        &(35.31±68.17         \\ \hline
7 - 8h & {Male}  & 316.59±93.26       & 225.82±82.13         & 224.89±71.30         & 215.82±75.55         \\ \hline
7 - 8h & {Female}  & 313.23±92.65        & 223.67±55.21         & 244.52±102.67        & 265.76±69.67         \\ \hline
\end{tabular}
}
\end{table*}

\section{Disussion}

We provide a comprehensive discussion of the study's findings, interpreting the effects of short-term aerobic exercise on executive function and exploring potential gender differences in response to varying exercise intensities. We also address the implications of these results within the broader context of cognitive neuroscience~\cite{luo2025intelligent,dong2024design,zheng2024triz,chen2024enhancing,sui2024application}, highlighting the mechanisms underlying the observed effects. Additionally, we discuss the limitations of the current study and suggest directions for future research, particularly regarding the generalizability of the findings to other age groups and the exploration of additional exercise variables. By synthesizing these insights, we aim to provide a deeper understanding of how exercise intensity, sleep quality, and gender may interact to influence cognitive performance, ultimately contributing to the development of more effective interventions to enhance executive function.

\subsection{Effects of short-term aerobic exercise intensity on executive function}

This study, focusing on middle-aged individuals, comprehensively reveals the effects of different intensities of short-duration aerobic exercise on various subcomponents of executive function, namely inhibition, shifting, and updating, under varying sleep quality conditions\cite{van2013structure,spaniol2022meta}. The findings indicate that short-duration aerobic exercise at moderate to high intensities significantly improves inhibition function, especially in individuals with shorter sleep durations. These results align with previous research; for instance, Hillman et al.\cite{hillman2009effect} found that moderate-intensity aerobic exercise notably enhances inhibition function in university students, similar to the findings of Sibley et al. and Ferris et al.\cite{sibley2006effects} using the Stroop task. Furthermore, this study demonstrates that aerobic exercise at low, moderate, and high intensities significantly improves shifting function, without significant differences between intensity levels. This is consistent with the findings of Netz et al.\cite{netz2007effect}, who showed that aerobic exercise improves shifting ability in older adults. Additionally, the study uniquely highlights the positive effects of aerobic exercise on updating function, with low-intensity exercise being the most effective, expanding the research on the relationship between aerobic exercise and executive function. The study proposes two potential mechanisms for these effects: the arousal level hypothesis, suggesting that short-duration aerobic exercise enhances metabolic activity and cerebral blood flow in executive function-related areas; and the catecholamine hypothesis, which posits that aerobic exercise increases neuroendocrine activity, improving executive function through changes in catecholamine neurotransmitters. While these hypotheses provide a plausible explanation, further research is needed to fully elucidate the underlying mechanisms. This research contributes experimental evidence to the "dose-response" effect of aerobic exercise on executive function and offers a practical foundation for developing exercise interventions based on aerobic exercise intensity to improve executive function.

\subsection{Gender differences in the effects of short-term aerobic exercise intensity on executive function}

Research on psychological differences between genders and the resulting advantages and disadvantages in various domains has long been a focal point in the field of psychology\cite{guimond2008psychological,wilkinson2022age}. Uncovering how aerobic exercise differentially impacts executive function across genders, and providing tailored education and intervention based on these findings, holds significant implications for improving executive function. This study reveals no significant gender differences in executive function, supporting the notion that such differences do not exist. Additionally, the findings indicate that short-term aerobic exercise of varying intensities impacts executive function in middle-aged individuals similarly across genders, highlighting, for the first time, the absence of gender differences in this context. Although the underlying reasons for the lack of gender-based differences are not yet fully understood, cognitive neuroscience research offers some insight. Advanced brain imaging studies have identified specific brain activity patterns associated with executive function, which are primarily regulated by age rather than gender\cite{bermejo2022gender,clayton2023psychological}. Given that participants in this study were healthy middle-aged adults with similar brain development levels, no gender differences were observed in executive function or its response to short-term aerobic exercise. These results suggest that when designing aerobic exercise interventions to improve executive function in middle-aged individuals, exercise intensity may be applied uniformly across genders\cite{otterbring2022pandemic}. However, the study only examines the impact of sleep quality and exercise intensity on executive function, and further research is needed to determine whether other variables, such as exercise type and duration, influence executive function or interact with gender\cite{pankowiak2023psychological}. Thus, it is premature to conclude that gender differences need not be considered when developing aerobic exercise interventions for executive function enhancement.

\subsection{Deficiencies and suggestions}
The limitations of this study should be acknowledged, as they highlight areas for improvement and further investigation. First, the participant pool consisted exclusively of middle-aged individuals, which raises concerns regarding the generalizability of the findings to other age groups. Cognitive functions can vary significantly across the lifespan, and factors such as age-related cognitive decline may influence the outcomes. Therefore, conclusions drawn from this study may not accurately reflect the effects of aerobic exercise on executive function in younger or older populations. Future research should incorporate a more diverse participant demographic to ensure that the findings can be applied more broadly across various age groups.

Second, the study primarily focused on the relationship between sleep duration, exercise intensity, and executive function, but it did not fully explore the potential "dose-response" effect of aerobic exercise. The dose-response relationship refers to how the effects of an intervention, in this case, aerobic exercise, can vary depending on the quantity and intensity of the exercise performed. By not examining other crucial components of aerobic exercise, such as exercise type, duration, and frequency, the current study may overlook significant interactions that could further elucidate how these factors collectively influence executive function.

To enhance the robustness of future research, it is essential to investigate these broader components of aerobic exercise and their interactions with exercise intensity. This approach will provide a more comprehensive understanding of the dose-response effect of aerobic exercise on executive function. Additionally, exploring how various types of aerobic activities, such as running, cycling, or swimming, impact cognitive performance may reveal insights into the most effective exercise modalities for enhancing executive function.

\section{Conclusion}

Research indicates that varying levels of sleep quality, combined with short-duration aerobic exercise of different intensities, selectively enhance the three dimensions of executive function—namely inhibition, updating, and shifting—in middle-aged adults. This effect has been observed to be consistent across genders, suggesting that both men and women can benefit from aerobic exercise in similar ways when it comes to cognitive functioning. The findings underscore the critical role that aerobic exercise can play as an effective intervention for improving executive function, highlighting its potential as a valuable tool in cognitive enhancement strategies.

Specifically, the selective enhancement of the three dimensions of executive function demonstrates the nuanced impact that both sleep and exercise can have on cognitive processes. Inhibition, which refers to the ability to suppress responses that are inappropriate or irrelevant, has shown marked improvement with higher intensity aerobic activities. Meanwhile, updating—related to the ability to refresh and maintain relevant information in working memory—also benefited from structured aerobic exercise, particularly when participants experienced adequate sleep quality. Lastly, shifting, which involves transitioning attention between different tasks or mental sets, was enhanced by the dynamic nature of aerobic workouts, further illustrating the adaptability of executive function in response to physical activity.

Moreover, this study lays a practical foundation for future interventions aimed at improving executive function by considering critical factors such as sleep duration, exercise intensity, and gender. Understanding the interplay between these elements can guide the design of tailored aerobic exercise programs that maximize cognitive benefits. For instance, future interventions could strategically incorporate varying exercise intensities based on individual sleep quality and adjust the duration of exercise sessions to better suit the needs of different age groups and genders.

Simultaneously, the results of this study have significant implications for the future of wearable technology and personalized health systems. By combining emotional and physiological data~\cite{huang2017all,lee2021neural,wu2022learning,komaromi2024enhancing,sun2023htec,wu2023jump,li2025decoupled,shojaee2025federated,ji2024nt,luo2023model}, future wearable devices can offer more comprehensive health monitoring, not only tracking physical activity but also assessing emotional states, stress levels, and more in real time, providing personalized health recommendations. For example, devices could automatically adjust exercise plans or offer relaxation suggestions based on emotional fluctuations. This data-driven approach to precise health management will drive the development of health tech, helping individuals better manage their physical and mental well-being while offering early warnings and personalized interventions for the healthcare industry.

For future improvements in this research, several key areas could be explored to enhance the overall impact and applicability of the findings. First, expanding the sample size and including participants from a broader age range could help determine whether the observed effects on executive function are consistent across different life stages. Additionally, incorporating longitudinal data would provide insights into the long-term benefits of aerobic exercise combined with sleep quality on cognitive performance. Another improvement would be the inclusion of more diverse exercise modalities, such as resistance training or mixed aerobic activities, to compare their effects on executive function in relation to sleep quality. Lastly, incorporating more objective measures of sleep quality, such as polysomnography or actigraphy, could provide a more precise understanding of how sleep influences cognitive outcomes, beyond self-reported data. By addressing these areas, future research could offer more robust and generalized conclusions, further optimizing interventions aimed at improving executive function and cognitive health.

\section*{Author Contributions}
All authors contributed significantly to this study. Yu Peng conceptualized the research, designed the methodology, supervised the project, and was involved in data collection and manuscript drafting. Guoqing Zhang led the statistical analysis and data interpretation, contributing to the research design and the development of the discussion and conclusion sections. Huadong Pang assisted with the experimental design and data collection, particularly focusing on the physical activity aspects, and played a key role in the manuscript's writing and technical accuracy review. All authors reviewed and approved the final manuscript.

\section*{conflicts of interest}
The authors declare that the research was conducted in the absence of any commercial or financial relationships that could be construed as a potential conflict of interest.

\bibliographystyle{cas-model2-names}
\bibliography{cas-refs}


\end{document}